\documentclass[preprint]{iucr}   
     \journalcode{S}              % Indicate the journal to which submitted
\usepackage{color}
\usepackage{amsmath}          % eqref   align
\usepackage{graphics}

\begin{document}

\title{Retrieval of the atomic displacements in the crystal from the coherent x-ray diffraction pattern}
\shorttitle{Retrieval of the atomic displacements in the crystal}

\cauthor[a]{A.~A.}{Minkevich}{aa.minkevich@gmail.com}{}
\author[a]{M.}{K\"{o}hl}
\author[b]{S.}{Escoubas}
\author[b]{O.}{Thomas}
\author[a,c]{T.}{Baumbach}

\aff[a]{ANKA/Institute for Photon Science and Synchrotron Radiation, Karlsruhe Institute of Technology, 76344 Eggenstein-Leopoldshafen, \country{Germany}}
\aff[b]{Aix-Marseille Universit\'{e}, CNRS, IM2NP, UMR 7334, 13397 Marseille Cedex 20, \country{France}}
\aff[c]{Laboratory for Applications of Synchrotron Radiation, Karlsruhe Institute of Technology, \country{Germany}}

\shortauthor{A.~A.~Minkevich et al.}


\maketitle 

\begin{abstract}
We investigate the retrieval of spatially resolved atomic displacements via the phases of the direct(real)-space image reconstructed from the strained crystal's coherent x-ray diffraction pattern. We demonstrate that limiting the spatial variation of the first and second order spatial displacement derivatives improves convergence of the iterative phase retrieval algorithm for displacements reconstructions to the true solution. Our approach is exploited to retrieve the displacement in a periodic array of silicon lines isolated by silicon dioxide filled trenches.
\end{abstract}

%\twocolumn

\newcommand{\todo}[1]{\textbf{\textcolor{red}{TODO: \underline{#1}}}}

\section{Introduction}

Strain engineering at the nanometer scale is a fast developing research direction attracting a lot of attention in the last decade. Properly engineered strains can enhance the mobility of charge carriers and hence the performances of electron devices (\citeasnoun{Ghani2003}, \citeasnoun{Signorello2013}). The control of the strain field in the channel of transistors require measuring strain with a nanometer resolution, which is a difficult challenge. X-ray diffraction is a very promising tool for this task because it is non destructive and has a high sensitivity to the strain (\citeasnoun{Baumbach04}). The high spatial resolution is currently only achievable by measurements in reciprocal space, since the sizes of the focussed x-ray beams are still far from the desired nanometer-dimensions (\citeasnoun{Stangl2009}).

In this paper we investigate the atomic displacement fields arising from elastic strains and the possibility to reconstruct them using x-ray coherent diffraction scattering. The approach based on the reconstruction of the coherent diffraction pattern - coherent diffractive imaging (CDI) - was already proved to be successful for the extraction of the shapes and electron density of some objects (\citeasnoun{Miao99}; \citeasnoun{Marchesini03}; \citeasnoun{Chapman06}; \citeasnoun{Shapiro05}; \citeasnoun{Rodenburg07}; \citeasnoun{Thibault08}; \citeasnoun{Nishino09}; \citeasnoun{Schroer_PRL}; \citeasnoun{Takahashi10}; \citeasnoun{Dierolf2010}). The ability of the x-ray CDI to reconstruct the local displacement field was demonstrated several years ago (\citeasnoun{Pfeifer06}; \citeasnoun{Robinson09}; \citeasnoun{Newton_2010}; \citeasnoun{Watari_2011}; \citeasnoun{Clark_2013}; \citeasnoun{Cha_2013}). The frequent failure to reconstruct the data sets from highly strained crystals (\citeasnoun{Minkevich08}; \citeasnoun{Diaz2010}) hindered, however, the wide application of the method. Recently the ptychography showed its ability to extend to the Bragg x-ray diffraction, offering the alternative way of overcoming the convergence problems (\citeasnoun{_Chamard2011}; \citeasnoun{Hruszkewycz_2012}; \citeasnoun{Takahashi_2013}). The improved convergence of the conventional method was also achieved by introducing the additional constraints requiring the electron density homogeneity and limiting the first spatial displacement derivatives (\citeasnoun{Minkevich08}). The successful application of this method, which was not possible otherwise, was illustrated for two particular sets of x-ray experimental data (\citeasnoun{Minkevich07}; \citeasnoun{Minkevich11PRB}). In the present manuscript we further improve the method to extend its applicability to the wider class of elastically strained crystals. In the next section we start the discussion from the influence of the choice of the origin of the reciprocal space image (RSI) on the results of strain field reconstruction. We discuss the relation between strain-induced broadening in reciprocal space and the inhomogeneity of the deformations. From this, we specify additional constraints limiting the phase gradients in direct space.

In this manuscript we focus on the benefit of direct space phase constraints without restricting chemical homogeneity. In particular, we investigate a possible strengthening of phase constraints by not only limiting the first order displacement derivatives (i.e., strain), but also the second order displacement derivatives (i.e., strain gradient). The implementation of the introduced constraints in the iterative algorithm is discussed in the last subsection of the Section II. Section III demonstrates the application of this approach to experimental x-ray scattering data\footnote{The demonstrated x-ray diffraction experiment is not a conventional CDI experiment in the direct sense of the word. This is the high resolution x-ray diffraction experiment with a wide partially coherent beam. Since the coherence of the incident beam is much larger than the single period of the structure, the scattering from the particular period is coherent. The experiment relies on the identity of the structure within individual periods.} of a periodically strained system consisting of oxide-filled trenches in crystalline silicon (\citeasnoun{Eberlein2008}). We expect applicability of the introduced additional constraints to a wide class of elastically strained nano- or micro- crystalline structures. 

\section{Coherent X-ray diffraction and inhomogeneous strain}

In case of crystalline objects the distribution of the x-ray scattered amplitude in reciprocal space around a reciprocal lattice point (RLP) can be approximated as 
\begin{equation} \label{diff1}
E(\vec{Q}) \sim \int {\rm d}^3\vec{r} \left[\varrho_{\vec{h}}(\vec{r}) {\rm
e}^{-{\rm i}\vec{h}.\vec{u}(\vec{r})}\right]{\rm e}^{-{\rm i}(\vec{Q}-\vec{h}).\vec{r}},
\end{equation}
where $\vec{h}$ and $\vec{Q}$ are the Bragg peak's reciprocal lattice vector and scattering vector, respectively. $\varrho_{\vec{h}}(\vec{r})$ is the Fourier component of the unstrained periodic electron density and  $\vec{u}(\vec{r})$ is the displacement field with respect to the non-strained lattice of bulk material. The formula (\ref{diff1}) assumes the validity of the first-order Born approximation. Moreover, it assumes that the strains are elastic and relatively small (\citeasnoun{Takagi69}) such as the crystal unit cells within the crystal volume are not deformed but only shifted as a whole, so that their scattering factors (corresponding to Fourier component of non-strained periodic electron density) are not affected by deformation. In addition, the application of the coherent imaging methods implies the crystal dimensions do not exceed the coherent volume of the incident x-ray radiation, which is presently in the order of several micrometers at the third generation synchrotron sources (\citeasnoun{Veen_2004}; \citeasnoun{Kohn_2000}). 

The expression (\ref{diff1}) expresses a Fourier transformation (FT) relation between x-rays' scattered amplitudes in reciprocal space ($E(\vec{Q})$) and the scatterer density distribution $\tilde{\varrho}(\vec{r})=\varrho_{\vec{h}}(\vec{r}) \exp(-{\rm i}\vec{h}.\vec{u}(\vec{r}))$ (i.e. direct space amplitude $\varrho_{\vec{h}}(\vec{r})$ and phase $\phi(\vec{r}) = \vec{h}.\vec{u}(\vec{r})$). Therefore, the inhomogeneously strained crystal under such conditions is recognized essentially as a phase object, whose phase brings the information about the atomic displacements. In case of chemical homogeneity it becomes a pure phase object. Once the phases of the scattered wavefield are known in far field, scatterer density distribution $\tilde{\varrho}(\vec{r})$ can be obtained by inverse FT. Since the phases are experimentally not accessible, the \emph{direct} investigation of the shape and strain of the crystalline nanostructure requires implementation of a phase retrieval method. The principle is based on 
an iterative loop of direct and inverse FT (towards the experimental intensity distribution and back to the sample space) and it may refine the genuine scatterer density distribution $\tilde{\varrho}(\vec{r})$ even by starting with a non realistic model (\citeasnoun{Fienup82}; \citeasnoun{Elser03}). The iterative method composes the core part of CDI. The well-known problems of the iterative reconstruction of the complex-valued direct-space objects (\citeasnoun{Minkevich08}; \citeasnoun{Koehl2013}) hinder the application of CDI to reliably study strain fields in highly strained crystals. The successful inversion of the phase objects representing the inhomogeneously strained crystals requires that appropriate additional constraints are introduced in the iterative scheme. This relies, besides the support constraint, on \textit{a priori} knowledge about the system to be reconstructed resulting in additional constraints during the iterative reconstruction process. First, we take a closer look on a iterative method having additional constraints limiting the allowed 
value of strain (phase gradient in direct space (DS)) \citeasnoun{Minkevich08}. The possible improvements of this iterative scheme and its extension to second order derivatives of the displacement are discussed in the successive sections.

\subsection{Strain-induced x-ray diffuse scattering and choice of the origin of RSI}

Phases of the reconstructed direct space density $\tilde{\varrho}(\vec{r})$ are highly sensitive to the choice of the origin of the RSI. This can be expressed as

\begin{eqnarray} \label{diff3}
E(\vec{Q}) \sim \int {\rm d}^3\vec{r} \left[\varrho_{\vec{h}}(\vec{r}) {\rm e}^{-{\rm i}(\vec{h}.\vec{u}(\vec{r}) +\Delta\vec{h}^{Bragg}.\vec{r})}\right] \nonumber\\
{\rm e}^{-{\rm i}(\vec{Q}-(\vec{h} +\Delta\vec{h}^{Bragg})).\vec{r}} = \nonumber\\
\int {\rm d}^3\vec{r} \left[\varrho_{\vec{h}}(\vec{r}) {\rm e}^{ -{\rm i} \phi(\vec{r}) }\right]
{\rm e}^{-{\rm i}(\vec{Q}-(\vec{h} +\Delta\vec{h}^{Bragg})).\vec{r}}.
\end{eqnarray}

\noindent Therefore any relocation of the origin of RSI from the original RLP is equivalent to the appearance of the corresponding linear term in the phase of $\tilde{\varrho}(\vec{r})$:

\begin{equation} \label{phases}
\phi(\vec{r})=\vec{h}.\vec{u}(\vec{r}) + \Delta\vec{h}^{Bragg}.\vec{r},
\end{equation}

\noindent where $\Delta\vec{h}^{Bragg}$ is a shift of the origin of the RSI from the RLP corresponding to the ideal reciprocal lattice denoted by $\vec{h}$.

The position of the RLP can be easily recognised for the substrate crystal by the presence of a strong diffraction peak in reciprocal space due to the strong Bragg reflection of the perfect crystal. In the case of a small deformed crystal the position of the original RLP cannot be determined so precisely without any reference, since the inhomogeneous strain modifies the diffraction pattern and makes it broader around RLP. It redistributes the scattering signal in a broader area around RLP, as compared with the scattering from the strain-free crystal, because of the local changes in the crystal lattice parameter. The not-slower-than-quadratic decay of intensities in reciprocal space which is usually observed for the unstrained crystal, is no more valid near the RLP. The width of the intensity cloud can be defined as $\Delta Q_p(\vec{h})$ along the particular $p$-direction (see, for example, $\Delta Q_z$ in Fig.~\ref{figSTI_RSM004_onlyGTRs} in the Section III), where $p$ is an arbitrary direction defined by 
the unit vector $\hat{e}_{p}$ so that $p = \hat{e}_{p}.\vec{p}$.

If the position of the RLP is not known, the uncertainties in the $\vec{h}$ position can be expressed in terms of unknown variation of $\Delta \vec{h}^{Bragg} = \tilde{\Delta} \vec{h}^{Bragg}$ (see (\ref{phases})). The determined strain field components (the displacement derivatives along particular $p$ axis), therefore, will differ from the true solution with respect to the substrate by the constant values $\frac{\tilde{\Delta} h^{Bragg}_p}{|\vec{h}|}$, where $\tilde{\Delta} h^{Bragg}_p$ is a $p$ component of $\tilde{\Delta} \vec{h}^{Bragg}$.

Large displacement gradients can often mask the shape effects in the RSI. The exception is a crystal truncation rod of the substrate crystal perpendicular to the crystal surface where the (slow) quadratic decay of intensities is observed. It is also possible to separate the effects of crystal shape and strain in the reciprocal space by considering several Bragg reflections. While crystal size broadening is independent of $\vec{h}$, strain broadening increases proportionally for larger $|\vec{h}|$ values. Therefore, $\Delta Q_p(\vec{h})$ could be more precisely identified by considering two different RSI having collinear reciprocal lattice vectors. Taking into account the shift property of the FT applied to the (\ref{diff1}), the strain-induced broadening $\Delta Q_p$ of the scattering signal around RLP in $p$-direction will be only possible if

\begin{equation} \label{broad}
\Delta Q_p(\vec{h}) \cong max\left[\frac{\partial \vec{h}.\vec{u}(\vec{r})}{\partial p}\right] - min\left[\frac{\partial \vec{h}.\vec{u}(\vec{r})}{\partial p}\right],
\end{equation}

\noindent here $\vec{r}$ can be represented as $(p, \vec{r}_{\perp})$, where $\vec{r}_{\perp}$ is a 2D vector representing two other orthogonal components of $\vec{r}$. Eq. (\ref{broad}) is a first order approximation based on the incoherent summation of the terms with different phase gradients in integral (\ref{diff1}). The values $\Delta Q_p^{+}(\vec{h}) \simeq max\left[\frac{\partial \vec{h}.\vec{u}(\vec{r})}{\partial p}\right]$ and $\Delta Q_p^{-}(\vec{h}) \simeq min\left[\frac{\partial \vec{h}.\vec{u}(\vec{r})}{\partial p}\right]$ describe the strain-induced broadening to lower and higher values from RLP along $p$ direction (see, for example, notes in Fig.~\ref{figSTI_RSM004_onlyGTRs}) and $\Delta Q_p(\vec{h}) = \Delta Q_p^{+}(\vec{h}) - \Delta Q_p^{-}(\vec{h})$.

\subsection{Constraints for the phase gradients}

Reconstruction of the diffraction pattern from an inhomogeneously strained crystal purely based on the geometry of the object (i.e. its shape) in direct space is not sufficiently robust for application to experimental data of samples with highly inhomogenous strain: The difficulty to reconstruct successfully depends on the value of inhomogeneity of the spatially resolved strain, which was demonstrated in the case of pure phase objects (\citeasnoun{Minkevich08}). The higher the inhomogeneity of the strain field, the higher the probability to meet the local minima and deep stagnation during the reconstruction (\citeasnoun{Minkevich08}; \citeasnoun{Koehl2013}). The presence of the strong inhomogeneity of the strain field can be recognized by the broadened distribution of the signal around RLP expressed by (\ref{broad}), which exceeds the spread of the corresponding crystal shape function decreasing not slower than $1/(\vec{Q} - \vec{h})^{2}$ in reciprocal space (RS). The dimensionless notation 

\begin{eqnarray}
\label{broad_dimless} \Delta_p u_j^{max} = max\left[\frac{\partial u_j}{\partial p}\right] - min\left[\frac{\partial u_j}{\partial p}\right],
\end{eqnarray}

\noindent of expression (\ref{broad}) is an important estimation of the complexity of the strained system to be reconstructed. It represents the strain field inhomogeneity along $p$ axis via maximum variation range of the corresponding displacement field derivatives. The phase retrieval method described in Ref.~\citeasnoun{Minkevich08} limits the phase variation of the direct-space object (and therefore the displacement derivatives) in order to satisfy the relation (\ref{broad_dimless}). This corresponds to the application of two inequalities constraints along $p$ axis:

\begin{equation} \label{constraint_phases1}
\Delta Q_p^{-}(\vec{h}) \leq \vec{h}.\frac{\partial \vec{u}(\vec{r})}{\partial p} \leq \Delta Q_p^{+}(\vec{h}),
\end{equation}

\noindent which is equivalent to Eq.(4) in Ref.~\citeasnoun{Minkevich08}. Taking derivative of $p$ variable from expression (\ref{phases}) results in

\begin{equation} \label{phases_derivative}
\frac{\partial \phi(\vec{r})}{\partial p} = \vec{h}.\frac{\partial \vec{u}(\vec{r})}{\partial p} + \Delta h^{Bragg}_p,
\end{equation}

\noindent where $\Delta h^{Bragg}_p$ is a $p$ component of $\Delta \vec{h}^{Bragg}$. Note that $\Delta h^{Bragg}_p \neq 0$ only if the reciprocal space origin differs from the RLP of the ideal lattice corresponding to the unstrained crystal. Now we can rewrite Eq.(\ref{constraint_phases1}) as

\begin{equation} \label{constraint_phases2}
\Delta Q_p^{-}(\vec{h}) + \Delta h^{Bragg}_p \leq \frac{\partial \phi(\vec{r})}{\partial p} \leq \Delta Q_p^{+}(\vec{h}) + \Delta h^{Bragg}_p.
\end{equation}

\noindent A shift of the origin of RS according to $\Delta h^{Bragg}_p = -\Delta Q_p^{-}(\vec{h})$ yields

\begin{equation} \label{constraint_phases_increase}
0 \leq \frac{\partial \phi(\vec{r})}{\partial p} \leq \Delta Q_p(\vec{h}),
\end{equation}

\noindent which corresponds to the direct-space phases to be increased along $p$-direction. Such a modification may be more convenient in implementation. One needs only to change the origin of RSI and take into account Eq.(\ref{phases}) when calculating the strain components after the image has been reconstructed. $\Delta h^{Bragg}_p$ is now an intentionally chosen shift of the origin of RSI along $p$-direction.

Similarly by setting $\vec{h}$ such as satisfying $\Delta h^{Bragg}_p = -\Delta Q_p^{+}(\vec{h})$ we ensure that the direct-space phase decreases along $p$-direction:

\begin{equation} \label{constraint_phases_decrease}
-\Delta Q_p(\vec{h}) \leq \frac{\partial \phi(\vec{r})}{\partial p} \leq 0.
\end{equation}

In the same way one can move the origin of RSI in the center of the strain-induced broadening along $p$ direction by introducing $\Delta \tilde{h}^{Bragg}_p$ satisfying $\frac{\Delta Q_p(\vec{h})}{2} = \Delta Q_p^{+}(\vec{h}) + \Delta \tilde{h}^{Bragg}_p$. Then, constraint (\ref{constraint_phases2}) can be rewritten as 

\begin{equation} \label{constraint_phases_middle}
\left| \frac{\partial \phi(\vec{r})}{\partial p} \right| \leq \frac{\Delta Q_p(\vec{h})}{2}.
\end{equation}

\noindent The notation (\ref{constraint_phases_middle}) was used in definition of phase constraints in Refs. \citeasnoun{Minkevich11EPL} and \citeasnoun{Minkevich11PRB}.

Adding these constraints to phase retrieval allowed successful reconstruction of the atomic displacements for two particular experimental data sets (\citeasnoun{Minkevich07}; \citeasnoun{Minkevich11EPL}) collected at European Synchrotron Radiation Facility (ESRF). These constraints to the phase gradients have been combined with constraints exploiting the spatial uniformity of the direct space amplitudes (resulting from the crystal's chemical homogeneity) which do not provide successful reconstructions for their own in case of higher strains (\citeasnoun{Minkevich08}; \citeasnoun{Koehl2013}). Despite the additional constraints to the phases, the reconstructions in \citeasnoun{Minkevich08} required permanent manual surveillance of the iterative solution and frequent revisions and modifications of the parameters. Usually, the iterative process starts with constraints (\ref{constraint_phases2}) exhibiting a phase limitation gap much smaller than $\Delta Q_p(\vec{h})$:

\begin{equation} \label{constraint_phases_tight}
n \Delta Q_p^{-}(\vec{h}) + b + \Delta h^{Bragg}_p \leq \frac{\partial \phi(\vec{r})}{\partial p} \leq n \Delta Q_p^{+}(\vec{h}) + b + \Delta h^{Bragg}_p,
\end{equation}

\noindent where $0 \leq n \leq 1$ and $b = 0$. The initial guess for the inversion of data from (\citeasnoun{Minkevich07}; \citeasnoun{Minkevich11EPL}) were usually twice tighter than their final value, i.e., $n \simeq 0.5$. We speculate, that these inversions have been possible because the displacement gradients of the investigated crystals satisfied Eq. (\ref{constraint_phases_tight}) in most parts of the crystalline volumes. The gradual relaxation of the constraint (\ref{constraint_phases_tight}) by shifting $n$ towards 1 was required with increasing number of iterations. This was done under surveillance until the constraint (\ref{constraint_phases_tight}) has relaxed to value in Eq. (\ref{constraint_phases2}).

We expect the method is the most functional if the information about the local distribution of the displacement gradients within the crystal volume is available. If the less inhomogeneously strained regions $\Omega_j$ can be identified, their local phase variation can be much more constrained than in the other parts where the strain is assumed to vary with higher magnitudes. This can be performed by specifying parameters $n = n_j$ and $b = b_j$ in (\ref{constraint_phases_tight}) applied in the crystal region $\Omega_j$. The parameter $b = b_j$ represents the magnitude of the average deviation between direct-space phase gradients in the subdomains $\Omega_j$ and must satisfy the initial inequalities (\ref{constraint_phases2}) by

\begin{equation}
(1 - n)\Delta Q_p^{-}(\vec{h}) \leq b \leq (1 - n) \Delta Q_p^{+}(\vec{h}). \nonumber
\end{equation}

Subdivision of the crystal volume in subvolumes $\Omega_j$ -- each with local phase constraints as tight as possible -- improves the chance to obtain the solution in a shorter time and with less efforts or provides the successful reconstruction for the cases where it is not be possible otherwise.

In practice, a nano-focused x-ray beam may help identifying the magnitude of the strain locally. By probing different parts $\Omega_j$ within the investigated crystal volume the corresponding diffraction patterns may be collected (\citeasnoun{Hanke08_APL}). The strain magnitudes may be estimated by investigating the collected images, in relation with the particular illuminated local crystal part $\Omega_j$. This information may be exploited for spatially resolved adjustment of the constraints (\ref{constraint_phases_tight}): an individual choice of $n_j$ and $b_j$ for each $\Omega_j$ part during phase retrieval. Note however, that if the domains $\Omega_j$ become too small, the size effects of the illuminated regions dominate and can hide strain broadening, complicating its estimation.

% Theoretically, if the regions $\Omega_j$ become very small and could approach the resolution in direct space (size of the pixel) the constraints (\ref{constraint_phases_tight}) applied to such fine elements $\Omega_j$ with specially chosen $n_j$ and $b_j$ limit locally also the second derivatives of the displacements $\frac{\partial^2 \vec{u}(\vec{r})}{\partial p^2}$.

\subsection{Constraints for the strain gradients}

To improve the convergence properties of the reconstruction algorithm with first order phase derivative constraints (described in the section above) further, we propose to limit components of strain gradients in addition, especially if such an information can be accessible. Thus, we have to limit the second derivatives of the direct-space phases

\begin{equation} \label{strains_derivative}
\frac{\partial^2 \phi(\vec{r})}{\partial p \partial s} = \vec{h}.\frac{\partial^2 \vec{u}(\vec{r})}{\partial p \partial s},
\end{equation}

\noindent imposing the limits

\begin{equation} \label{constraint_strains}
\Delta S_{ps}^{-}(\vec{h}) \leq \frac{\partial^2 \phi(\vec{r})}{\partial p \partial s} \leq \Delta S_{ps}^{+}(\vec{h}),
\end{equation}

\noindent where $p$ and $s$ are two axes over which the differentiation is performed ($p$ and $s$ may coincide). $\Delta S_{ps}^{-}(\vec{h})$ and $\Delta S_{ps}^{+}(\vec{h})$ are the corresponding lower and upper limits. These constraints can be applied together with the constraints limiting the first phase derivatives variation (\ref{constraint_phases2}), and therefore may strengthen them. Detailed investigations of realistic strained crystal models (e.g. by finite element modelling) reveals the small magnitudes of $\frac{\partial^2 \vec{u}(\vec{r})}{\partial p \partial s}$ in most of the crystal's volume. Up to now, we do not have a good estimator for the maximum values of strain gradient components directly from the measured diffraction pattern. However, the limitations might be taken from already known data for other systems of same materials and assuming a similar elastic behavior of the material under similar stresses.

The constraint (\ref{constraint_strains}) could impose tighter limitations on the strain gradients and relax them gradually with iterations (similarly to (\ref{constraint_phases_tight}) by changing $n$ with $b=0$). Once the constraint, being narrow enough, will be satisfied within a considerable part of the crystal volume, we expect, that the chance to reconstruct the object successfully is substantially increased. The reconstructions should always reveal the same strain distribution for different magnitude limitations in order to distinguish them from the results of stagnation in the local minima. 

Moreover, the constraints are suitable for application in part of the crystalline volume only. This might be required at material interfaces between constituent materials where the second order phase derivatives have discontinuities. This imposes additional knowledge about the internal geometry of the investigated crystalline structure.

\subsection{Implementation of additional constraints in phase retrieval algorithm}

The direct-space additional constraints can be computationally implemented in different ways both for the homogeneity of the amplitudes and for the limit on the phase variation. Here we describe one of the possibilities of implementation of these constraints in a very general way. The approach allows imposing limitations on strain as well as strain gradients depending on the definition of the corresponding parameter set. It is based on the modification of the hybrid input-output (HIO) algorithm introduced by Millane (\citeasnoun{Fienup82}; \citeasnoun{Millane97}). If $g_k(i)$ and $g'_k(i)$ are the complex values correspondingly of the input and output of a point $i$ of the direct-space object at $k$ iteration (for definitions of input and output of HIO algorithm see Ref.~\citeasnoun{Fienup82}), then the input of the next iteration $k+1$ of our modified algorithm HIO$^{AP}$ containing the additional constraints is defined as

%\onecolumn

\begin{subequations}
\begin{align} 
\label{HIO_ap:a} 
g_{k+1}(i) = \left \{ \begin{array}{ll}
                       g'_{k}(i), & \left| |g'_{k}(i)|-|c_{k}(i)| \right|\le\epsilon_{a} \quad \& \quad \left| \phi_{i}-\psi_{i} \right|\le\epsilon_{p} \\
                       g_{k}(i)+\beta (c_{k}(i)-g'_{k}(i)), & \left| |g'_{k}(i)|-|c_{k}(i)| \right|> \epsilon_{a} \quad \| \quad \left| \phi_{i}-\psi_{i} \right|> \epsilon_{p} 
                      \end{array} \right., \\
\label{HIO_ap:b} 
|c_{k}(i)|=\left \{ \begin{array}{ll}
                     0, & i \notin \gamma \\
                     \frac{1}{N_{V_a(i)}} \sum_{j \in V_a(i)} |g'_{k}(j)|, & i \in \gamma
                    \end{array} \right., \\
\label{HIO_ap:c} 
\psi_{i}= \left \{ \begin{array}{ll}
                     \phi_{i}, & i \notin \gamma \\
                     \Phi + \frac{1}{N_{V_p(i)}} \sum_{j \in V_p(i)} \arg_{\mathrm{unwrap}}(g'_{k}(j)), & i \in \gamma
                    \end{array} \right., \\
\nonumber
c_{k}(i)=|c_{k}(i)|e^{\psi_{i}}, \\
\nonumber
\quad \phi_{i} = \arg(g'_{k}(i)),
\end{align}
\end{subequations}

%\twocolumn

\noindent where $\beta$ is a parameter taken usually from the interval [0.5, 1.0] (see definition of HIO algorithm \citeasnoun{Fienup82}). $\gamma$ is the support area in direct space, i.e, the shape of the direct-space object. $\epsilon_a$ and $\epsilon_p$ are the thresholds for applying the constraints \eqref{HIO_ap:a}--\eqref{HIO_ap:c} to each individual point $i$ in the direct-space image (DSI) for amplitudes and phases correspondingly. $V_a(i)$ and $V_p(i)$ define the vicinities (the set of neighbouring points) of point $i$, $N_{V_a(i)}$ and $N_{V_p(i)}$ are the numbers of points in the corresponding vicinities $V_a(i)$ and $V_p(i)$. The amplitudes and phases of $c_k(i)$ are independently calculated during each iteration $k$ by (\ref{HIO_ap:b}) and (\ref{HIO_ap:c}) correspondingly. $\Phi$ is a compensation parameter ensuring the equilibrium between average value $\psi_{i}$ and $\phi_{i}$ in (\ref{HIO_ap:a}) (see its definition below). The $\arg()$ operator takes the phase from its operand, whereas the $\arg_{\mathrm{unwrap}}()$ 
operator requires local unwrapping of the phases with respect to the neighbours before summation. The unwrapping can be done easily as long as the assumption holds that the difference between the unwrapped phase of adjacent pixels is always less than $\pi$. This is guaranteed by the direct-space sampling requirements in order to extract unambiguously the atomic displacements from the reconstructed phases in direct space.

In the case of a chemically homogeneous crystalline part $\gamma_{hom} \subseteq \gamma$, $V_a(i)$ may be defined as the set of all points in $\gamma_{hom}$ for all points from $i \in \gamma_{hom}$. In this case, $|c_{k}(i)|$ does not depend on $i$ for all points  $i \in \gamma_{hom}$. This significantly speeds up numerical computations of the modified HIO algorithm.

By the specific definition of the vicinity $V_p(i)$, we choose the particular phase constraints: Additional constraints to the phase gradients along the particular direction $p$ can be introduced in \eqref{HIO_ap:a}--\eqref{HIO_ap:c} by introducing only one neighbouring to point $i$ along $p$ direction to $V_p(i)$. Then, (\ref{HIO_ap:a}) limits the maximum allowed difference between phases in two neighbouring points, which corresponds to the maximally allowed phase gradient. If the origin of RSI coincides with the center of the strain-induced coherent diffuse scattering cloud along $p$ direction, the compensation parameter $\Phi = 0$ in (\ref{HIO_ap:c}) and the constraints to the phases in (\ref{HIO_ap:a}) are introduced in the same way as in (\ref{constraint_phases_middle}). Thus, $\epsilon_p$ is equivalent to the maximum allowed discrete phase derivative via

\begin{equation} \label{epsilon_p_relation1}
\epsilon_p \sim \Delta p \left|\frac{\Delta \phi(i)}{\Delta p}\right|,
\end{equation}

\noindent where $i$ is any point in direct space, $\Delta p$ is a step along the $p$ direction of DSI and $\frac{\Delta \phi(i)}{\Delta p}$ is a direct space discrete phase derivative. Shifting the origin of RS to another position leads to the requirement of having non-zero asymmetry compensation parameter $\Phi$ in (\ref{HIO_ap:c}). In this case the allowed phases difference for the discrete data points inside the support area $\gamma$ is

\begin{equation} \label{asym_phase_conditions_HIO_ap}
\epsilon_p^{min} \le \Delta \phi(i) \le \epsilon_p^{max}.
\end{equation}

\noindent$\Delta \phi(i) = \phi(i) - \phi(i-1)_{\mathrm{unwrap}}$ is the unwrapped phase difference between two neighbouring points along $p$-direction. For example, if the RS origin coincides with the lower edge of the strain-induced cloud in RS, the constraints to the phases coincide with Eq.~\eqref{constraint_phases_decrease}. For discrete data,  Eq.~\eqref{constraint_phases_decrease} is equivalent to (\ref{asym_phase_conditions_HIO_ap}) with $\epsilon_p^{min} \simeq -\Delta Q_p(\vec{h}).\Delta p$ and $\epsilon_p^{max} = 0$.

This asymmetry between $\phi_{i}$ and $\psi_{i}$ inside $\gamma$ for $V_p(i)$ containing only one previous index along $p$-direction is compensated by introducing a linear compensation parameter $\Phi = \frac{\epsilon_p^{max} + \epsilon_p^{min}}{2}$. The threshold to phases in (\ref{HIO_ap:a}) is expressed as $\epsilon_p = \frac{\epsilon_p^{max} - \epsilon_p^{min}}{2}$. Thus, the phase constraints in \eqref{HIO_ap:a}--\eqref{HIO_ap:c} are independent of the choice of the origin in RSI.

Introducing the several neighbouring points to $V_p(i)$ corresponds to limiting the variation of higher order phase derivatives when applying \eqref{HIO_ap:a}--\eqref{HIO_ap:c}. The application of \eqref{HIO_ap:a}--\eqref{HIO_ap:c} with $V_p(i)$ consisting of two neighbouring points of $i$ ($i$ itself is not included) along particular $p$ direction results in limiting the second order phase derivative (strain gradients). This can be seen from the definition of the discrete approximation of the second phase derivative 

\begin{equation} \label{second_derivative}
\notag
\frac{\Delta^2 \phi(i)}{\Delta p^2} = \frac{\phi(i-1) - 2\phi(i) + \phi(i+1)}{\Delta p^2}
\end{equation}

\noindent in an arbitrary point $i$ along $p$-direction and the particular relations with elements of \eqref{HIO_ap:a}--\eqref{HIO_ap:c} $\arg(g'_{k}(i)) \sim \phi(i)$ and $\arg(c_{k}(i)) \sim \frac{\phi(i-1) + \phi(i+1)}{2}$ inside $\gamma$ (the compensation parameter $\Phi = 0$). Correspondingly, $\epsilon_p$ in (\ref{HIO_ap:a}) defines the allowed discrepancy from  a vanishing  strain gradient and is related to the maximum allowed discrete second order phase derivative along $p$ direction via

\begin{equation} \label{epsilon_p_relation2}
\epsilon_p \sim \frac{\Delta p^2}{2} \left|\frac{\Delta^2 \phi(i)}{\Delta p^2}\right|.
\end{equation}

\noindent The relation between second order phase derivatives and second order displacement projections is expressed via Eq.~(\ref{strains_derivative}).

\section{Displacement reconstruction in stressed Silicon lines}

\subsection{Etched Silicon lines and X-ray diffraction measurements}

We focus here on a particularly relevant example of strained crystal. We have performed the analysis of spatial strain distribution of silicon-based semiconductor nanostructures using a combination of High Resolution X-ray Diffraction and the proposed extension of current phase retrieval methods. The periodic nanostructure is created using Shallow Trench Isolation (STI) technology, which is used in many microelectronics applications like non-volatile memories (\citeasnoun{Senez2001}). Deep trenches are etched in the silicon (001) substrate by photolithography and filled with SiO$_2$ in order to isolate the memory cells electrically. The complex production process -- consisting of many different steps -- generates high mechanical stresses which may damage the device and reduce its reliability (\citeasnoun{Arzt91}, \citeasnoun{Sauter92}). Moreover the strain induced in the silicon crystal in-between the trenches modifies the band structure and hence the mobility of charge carriers. This justifies the importance of obtaining the locally resolved strain 
information in such devices.

\begin{figure} \label{figSTI_TEM}               %Fig.
\resizebox{8.5cm}{8.5 cm}{\includegraphics{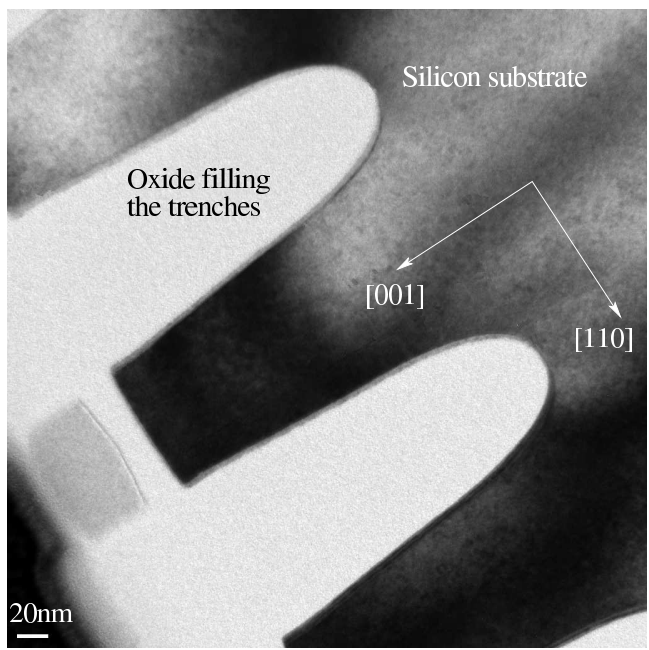}} 
\caption{Transmission Electron Microscopy image of the STI etched Silicon lines, with trenches filled with SiO$_2$. The normal to the grating surface corresponds to the [001] crystallographic direction. The Silicon periodic grating structure is fully covered by Silicon oxide.}
\end{figure}

In particular, we investigate stressed STI silicon lines with a periodicity of 200 nm. The trenches are filled with SiO$_2$ and oriented along the [1$\bar{1}$0] direction. Their depth is 250 nm. The crystalline silicon line active area on top of the silicon part of the nanostructure is approximately 90 nm wide. A Transmission Electron Microscopy (TEM) cross-section image (see Fig.~\ref{figSTI_TEM}) shows that side walls of the trench are not vertical. This is done intentionally during production in order to avoid sharp edges resulting in stress singularities in silicon, which could induce dislocation generation.

\begin{figure} \label{figSTI_RSM004}               %Fig.
\resizebox{8.5cm}{4.5 cm}{\includegraphics{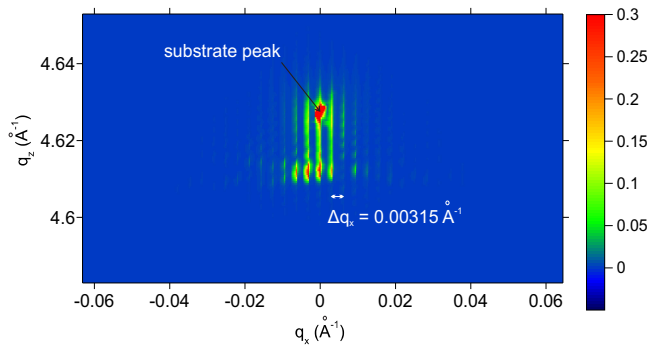}}
\caption{(Color online) High resolution x-ray diffraction measurements of the STI etched silicon lines in the vicinity of the 004 Bragg reflection of silicon. The intensities are plotted in logarithmic scale. For better visibility the color bar covers only the diffuse scattering intensities from 0 to 0.3 interval, although the substrate peak is very high and reaches a magnitude of 4.75.}
\end{figure}

The diffraction measurements from these STI lines were performed at BM32 beamline at the ESRF. A monochromatic x-ray beam with an energy of 9 keV and 0.2 $\times$ 0.2 mm$^2$ transverse cross-section impinges the specimen with an inclination near the 004 Bragg angle. High resolution reciprocal-space images in the vicinity of the 004 RLP was measured by use of a Si(111) triple crystal analyzer, placed one meter away downstream the specimen. The measurements were performed point by point by rocking the specimen in the vicinity of 004 reflection and scanning the Ewald sphere by changing the $2\theta$ angle (and consequently the exit angle). The RSI is shown in Fig.~\ref{figSTI_RSM004}. The measurements were performed in the plane perpendicular to the specimen surface and to the trenches direction. The periodicity of the sample gratings results in the appearance of the periodic grating maxima in the horizontal direction of the RSI (\citeasnoun{Minkevich11PRB}) as can be seen in Fig.~\ref{figSTI_RSM004}. They are well resolved, proving the high in-plane 
coherence of the incoming x-ray beam and the perfection of the periodic structure. The experiment can be considered as a pseudo coherent diffraction, being a partially coherent, it provides the coherent scattering from a single STI line. The periodicity of the structure allows considerable magnification of the scattering signal (\citeasnoun{Minkevich11PRB}) which is highly concentrated on the grating satellite maxima. The distances between the grating maxima, therefore, define the sampling step in the horizontal direction (highlighted in Fig.~\ref{figSTI_RSM004}) in reciprocal space. Its inverse distance corresponds to one grating period in direct space. The step in the vertical direction is controlled during the measurements and chosen to satisfy the oversampling requirements for our structure. The perfect crystal's substrate underneath is responsible for strong scattering in the close vicinity of 004 RLP. It violates the FT relation with direct space, and thus, we cannot account for the peak data during our reconstructions. The essential contribution to the signal in reciprocal space in other parts of the measurement originates from the strained part of the silicon 
nanostructure. Therefore, the sampling in reciprocal space should be chosen fine enough in order to provide enough data for the reconstruction of the strained part of the silicon. The chosen sampling in reciprocal space corresponds to the direct-space image of one period of 200 nm wide and to the 1250 nm height, which is assumed to satisfy the oversampling requirements for the corresponding strained part of the crystal (\citeasnoun{Miao98}).

\subsection{Phase retrieval in STI lines}

\begin{figure} \label{figSTI_RSM004_onlyGTRs}               %Fig.
\resizebox{8.5cm}{4.5 cm}{\includegraphics{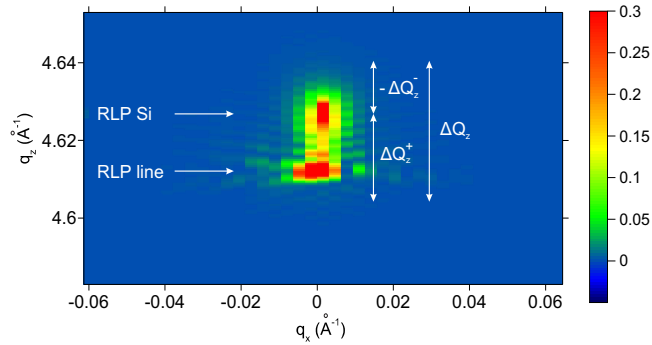}}
\caption{(Color online) Reciprocal space image extracted from the diffraction measurements (see Fig.\ref{figSTI_RSM004}) by taking the intensities only from the periodic satellite maxima and subtracting the noise background.  The color bar is logarithmic and covers only the diffuse scattering intensities from 0 to 0.3 interval as in Fig.\ref{figSTI_RSM004}. The dimensions of the created image are 141 (vertically) by 41 (horizontally) pixels.}
\end{figure}

The reconstruction of the atomic displacement field in a STI line requires preprocessing of the corresponding RSI: only the scattering data from the grating satellite maxima is used. A spatially varying background (because of the strong substrate scattering) is estimated as minimal magnitudes between the periodic maxima. These values are subtracted locally from the neighbouring grating maxima in RSI, which are then extracted for the further analysis. The resulting image is shown in Fig.~\ref{figSTI_RSM004_onlyGTRs}. The whole region shown in Fig.~\ref{figSTI_RSM004_onlyGTRs} is used for the analysis. The diffraction pattern from Fig.~\ref{figSTI_RSM004_onlyGTRs} (or the same from Fig.~\ref{figSTI_RSM004}) exhibits two well defined maxima. The strongest one is the scattering from the substrate and it corresponds to the 004 RLP of silicon. The appearence of the second maxima is connected with the uniform change of the crystal lattice parameter in the most part of the silicon line. Due to side pressure from the trench-filled oxide, the line undergoes horizontal compression, resulting in the vertical stretching of the mean lattice parameter in the line, which is revealed by the 
measurements in the vicinity of 004 RLP. The investigation of the lateral displacement components in a STI line requires measuring of an asymmetrical RLP.

The phase retrieval algorithm with and without additional constraints \eqref{HIO_ap:a}--\eqref{HIO_ap:c} was applied to the two-dimensional (2D) RSI from Fig.~\ref{figSTI_RSM004_onlyGTRs}. The support in direct space was approximated from the TEM image (see Fig.~\ref{figSTI_TEM}) and corresponds to the 2D cross-section of the shape of the strained silicon line. The support lower edge was chosen to lie about 200 nm deeper beyond the trench bottom, and it should be enough to restore the deformation field, which is assumed to be contained inside the defined support area. The oversampling ratio in this case is about 4.

Six pixels on the main truncation rod in the close vicinity of the RLP of silicon (strong substrate peak) were not constrained during the reconstruction process and were allowed to vary freely. The convergence process is monitored by the level of the metric error

\begin{equation} \label{error_metric}
E^2_k = {\sum^{N}_{i=1} \left( \left|F^{\mathrm{calc}}_i\right| - \sqrt{I^{\mathrm{meas}}_i} \right)^2}/{\sum^{N}_{i=1} I^{\mathrm{meas}}_i},
\end{equation}

\noindent where $|F^{\mathrm{calc}}_i|$ is the magnitude of the calculated amplitude and $I^{\mathrm{meas}}_i$ is the measured intensity of point $i$ in the RSI from Fig.~\ref{figSTI_RSM004_onlyGTRs}.

\begin{figure} \label{figSTI_result_wrong}               %Fig.
\resizebox{8.5cm}{8.5 cm}{\includegraphics{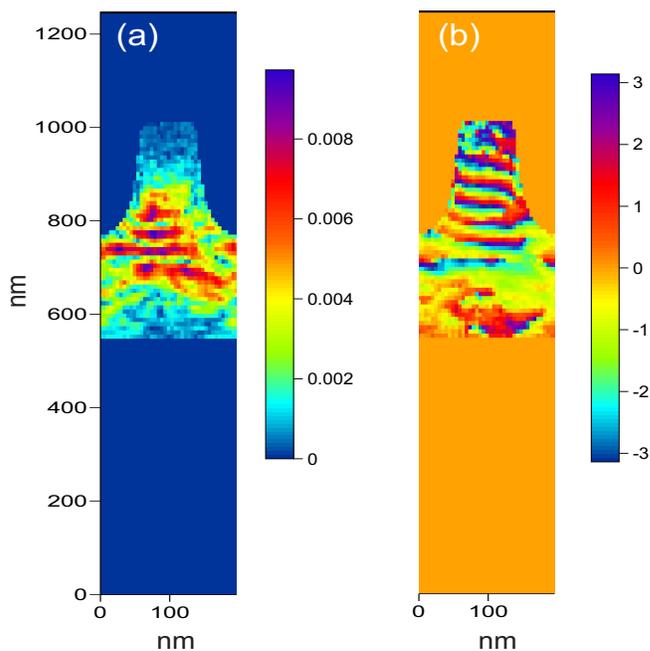}}
\caption{(Color online) A typical result for the displacement field z-component of our STI line based on the  HIO+O$_R$-algorithm and the RSI shown in Fig.~\ref{figSTI_RSM004_onlyGTRs}: (a) amplitudes and (b) phases.}
\end{figure}

The recently developed extension of the hybrid input-output (HIO) algorithm (\citeasnoun{Fienup82}) based on randomized overrelaxation (HIO+O$_R$) \citeasnoun{Koehl2012} was applied to the RSI from Fig.~\ref{figSTI_RSM004_onlyGTRs}. The algorithm does not rely on any additional constraints beyond shape $\gamma$ and ${I^{\mathrm{meas}}_i}$. Moreover, it was found to be more efficient than classical HIO+ER, if  applied to diffraction patterns of strained nanostructures (\citeasnoun{Koehl2013}). The method extends its successful applicability to larger values of strain inhomogeneity (\ref{broad_dimless}) where the classical HIO+ER algorithm fails (\citeasnoun{Koehl2013}). It was, however, shown that at some strain inhomogeneity (\ref{broad_dimless}), even the HIO+O$_R$ algorithm fails to reconstruct the displacement field (\citeasnoun{Koehl2013}; \citeasnoun{Minkevich08}). A typical result of the HIO+O$_R$ algorithm -- combined with ER -- to the data from Fig.~\ref{figSTI_RSM004_onlyGTRs} is shown in Fig.~\ref{figSTI_result_wrong}. Hence, despite the 
benefits provided by HIO+O$_R$, the algorithm cannot reconstruct the strained silicon line. 

The next step was an application of the phase retrieval algorithm HIO$^{AP}$ based on the additional constraints \eqref{HIO_ap:a}--\eqref{HIO_ap:c}. The usual way of iterative reconstruction involves iterative cycle repetitions, where each cycle contains several iterations of HIO$^{AP}$ (or HIO+O$_R$ in previous case) followed by several iterations of ER. We used an algorithm cycle consisting of 50 iterations of HIO$^{AP}$ followed by 20 iterations of ER. Several different parameters sets were used to study the convergence of the full algorithm with the additional constraints \eqref{HIO_ap:a}--\eqref{HIO_ap:c}. Fixing the particular set of parameters in our HIO$^{AP}$ is not always the best strategy. Human monitoring and interaction with the iterative process by changing some of the algorithm's parameters (in a similar way as introduced in (\ref{constraint_phases_tight}) by changing $n$) was found to be often useful in order to improve the convergence of the algorithm to the solution for the data set. A dedicated software with graphical user 
interface was developed in order to monitor and interact with the convergence process.

The constraints on the first phase derivatives along vertical $z$ as well as horizontal $x$ direction were introduced by HIO$^{AP}$. In this case $V_p(i)$ consists of one neighbouring point along $z$ or $x$ for each point $i$. $\epsilon_p$ limits the phase derivatives according to (\ref{epsilon_p_relation1}). The freedom in the choice of the origin of RSI is compensated by the corresponding value of $\Phi$ in (\ref{HIO_ap:c}), which corresponds to the shift of origin from the center of the strain-induced broadening in RS. The sampling of our DSI, which depends on the dimensions of the measured area in reciprocal space, is $\Delta x= 4.9$ nm in horizontal direction and $\Delta z = 8.9$ nm in vertical direction. The natural direction for phase gradient limitations is $z$ because the strain-induced broadening $\Delta Q_z$ is larger (see Fig.~\ref{figSTI_RSM004_onlyGTRs}) and consiquently the variation freedom of the displacement gradients is higher.

Constraints on the phase variation in the vertical direction $z$ --- by introducing the magnitude of $\epsilon_p = \frac{\Delta Q_z.\Delta z}{2} \approx 1.5$ (see in Fig.~\ref{figSTI_RSM004_onlyGTRs}) --- does not help inverting our data in the framework of the HIO$^{AP}$ algorithm: several tens of trials having thousands of iterations did not provide the solution. 

Moreover, the solution was never reached if both directions ($z$ or $x$) have been constraint by $\epsilon_p$ equal to 0.2, 0.5, 1 and 2. In all sets of parameters we used a relatively large $\epsilon_a = 0.00125$, which is approximately a 30\% of average amplitude of the solution. We avoided manual adjustment of the constraints (see (\ref{constraint_phases_tight}) and discussion) throughout the iterative process, since it relies on intuition and cannot be generalized easily to other structures. Instead, we introduced additional knowledge about the localization of the strain in the substrate and in the line regions.

\begin{figure} \label{figSTI_result_firstDeriv}               %Fig.
\resizebox{8.5cm}{8.5 cm}{\includegraphics{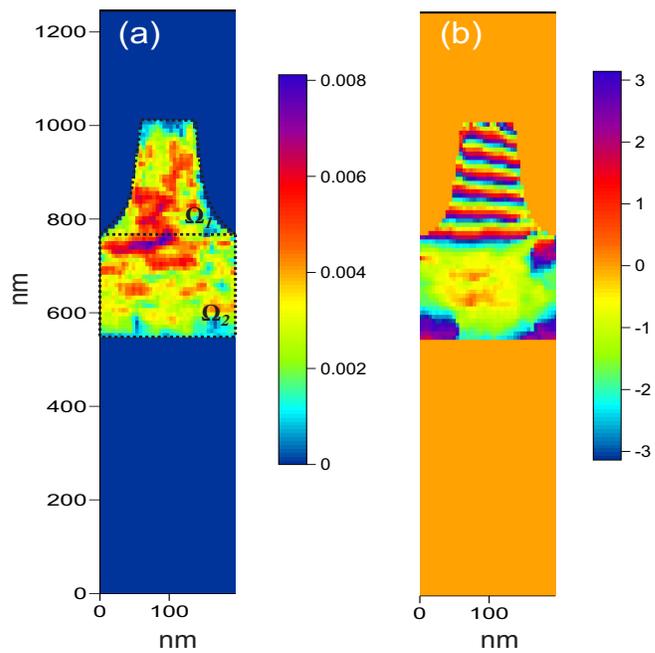}}
\caption{(Color online) The reconstructed direct-space image of one period of the STI periodic silicon structure: reconstructed (a) amplitudes and (b) phases. The convergence is achieved by limiting the first vertical displacement derivatives locally in $\Omega_1$ and $\Omega_2$ regions.}
\end{figure}

The support area was subdivided into two regions, namely, the etched line itself ($\Omega_1$) and the region underneath ($\Omega_2$), which can be seen in Fig.~\ref{figSTI_result_firstDeriv}(a). Each of them is responsible for the scattering in the vicinity of RLP of line and silicon substrate correspondingly, which are separated in RS (see Fig.~\ref{figSTI_RSM004_onlyGTRs}). The parameters of HIO$^{AP}$ algorithm were chosen differently for $\Omega_1$ and $\Omega_2$, taking into account the local broadening and the shift of both peaks. A successful choice is the vicinity $V_p(i)$ consisting of one previous point along $z$ and corresponding to the limiting of the first vertical phase derivatives. The corresponding parameters of \eqref{HIO_ap:a}--\eqref{HIO_ap:c} differ for both support regions $\Omega_1$ and $\Omega_2$. For $\Omega_2$ $\epsilon_{p} = 0.45$ with $\Phi = 0$, if the origin of RSI coincides with RLP of silicon substrate. For $\Omega_1$ $\epsilon_{p} = 0.335$ with $\Phi = 1.335$ (the minus coming from the definition 
of FT is not shown here). The related solution, however, contains some artifacts resulting from the subdivision of the system in two parts (see the typical example of the solution in Fig.~\ref{figSTI_result_firstDeriv}). For this system we found a more reliable approach based on limiting the second phase derivatives, which allows for the successful inversion.

\begin{figure} \label{figSTI_result}              
\resizebox{8.5cm}{8.5 cm}{\includegraphics{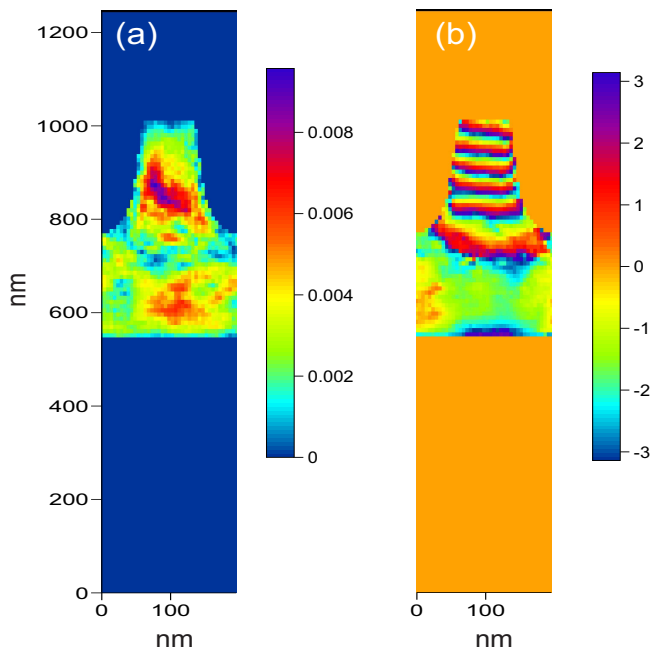}}
\caption{(Color online) Full reconstructed direct-space image of one period of the STI periodic silicon structure: reconstructed (a) amplitudes and (b) phases. The image is obtained through the inversion of the prepared diffraction pattern from Fig.\ref{figSTI_RSM004_onlyGTRs}.}
\end{figure}

Constraints on the second order phase derivatives (by defining $V_p(i)$ as two neighbouring points of point $i$ in \eqref{HIO_ap:a}--\eqref{HIO_ap:c}) were then introduced. Along the horizontal direction ($p = x$) the algorithm did not show the desired robustness for the inversion of our system. Apart from a few seldom convergences to an image resembling the main solution features, the algorithm usually ended up with different results which are far from the solution. These rare cases have only been observed for very small values of $\epsilon_p$, namely 0.02 and 0.05 which is related to the maximum allowed second phase derivative via Eq.~(\ref{epsilon_p_relation2}). Limiting the strain gradients in lateral direction does not have a direct impact on vertical phase gradients, which large variation freedom (see (\ref{broad_dimless})) is the main reason of the algorithm stagnation. 

\begin{figure} \label{figSTI_displ}
\resizebox{8.5cm}{8.5 cm}{\includegraphics{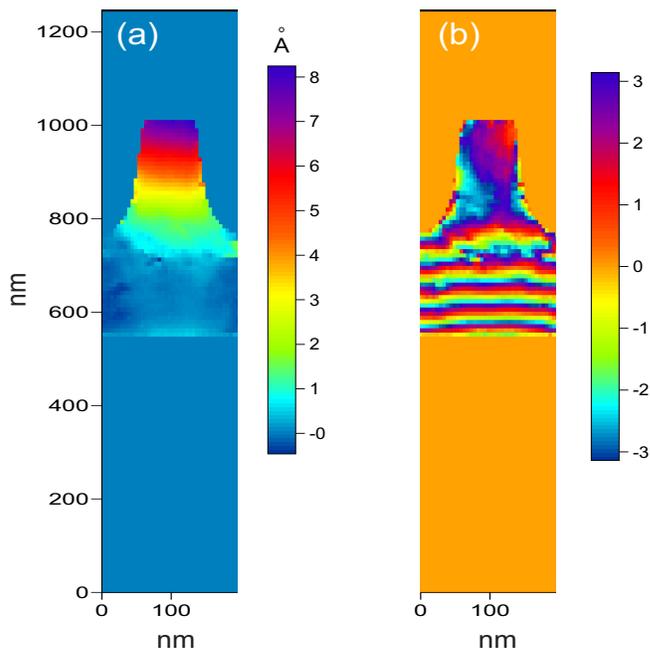}}
%\resizebox{8.5cm}{8.5 cm}{\includegraphics{figSTI_displacement_contour.eps}}
\caption{(Color online) (a) vertical component of displacements in the STI line calculated from the reconstructed direct-space phases shown in Fig.~\ref{figSTI_result}(b),  (b) direct-space phase image corresponding to the phases shown in Fig.~\ref{figSTI_result}(b), but the origin of RSI coincides with the RLP of line.}
\end{figure}

The successful choice was the limitation of phase variation along the vertical direction ($p = z$). The strong spatial variations of the vertical phase gradients, in case of reconstruction of wrong results (such as e.g. in Fig.~\ref{figSTI_result_wrong}), appear not in the correct places of the object. The splashes of the magnitudes of the second phase derivatives, which do not exist in case of the correct solution, appear to compensate such wrong phase behavior and are limited now by the current choice of the algorithm parameters. The level of metric error (\ref{error_metric}) is a complementary measure of the successful convergence together with the visual evaluation of the results, since the presence of noise and other experimental artifacts make only small differences between metric errors of local minima and the correct solution. After several cycles the metric error usually approaches the level of 0.01 - 0.05. The solution is usually found once the metric error reaches 0.004 - 0.007. Therefore, the drop in the metric error after approaching the solution is very often relatively small. The chance to reach the solution was best for $\epsilon_p = 0.02, 0.05$. For larger values ($\epsilon_p = 0.1$ and $0.2$) it was also possible to find the solution, the convergence to the correct result was, however, less stable. For even larger values, successful reconstructions have been observed very seldom. If the metric error does not drop below 
0.01 level after several thousands of iterations, it indicates stagnation which cannot be easily overcome by the current set of constraints parameters \eqref{HIO_ap:a}--\eqref{HIO_ap:c}. Therefore, a new trial -- having new random phases in reciprocal space as initial guess -- is taken and the procedure is repeated. An interesting point here is that we barely needed to interactively adjust the value of $\epsilon_p$ throughout the iterative process: It was only required for later refinement aiming to decrease the error to a lower level. At that point, the main features of the phase field of the solution were already reconstructed.

A typical result for the solution is shown in Fig.~\ref{figSTI_result}. The non-perfectly homogeneous distribution of amplitudes within the silicon line volume might be the sign of the presence of the disturbances violating the validity of the model, such as, for example, the strong scattering from the substrate. The inhomogeneities of the amplitudes were observed when reconstructing the blurred diffraction data (\citeasnoun{Vartanyants01_coherence}). The quality of the measured data in terms photon noise and signal to noise ratio can also affect the quality of reconstructions (see e.g. \citeasnoun{Koehl2013}). 

The phases clearly reflect an almost linear increase along the $z$-direction inside the line, which corresponds to a uniformly changed vertical lattice parameter. The corresponding vertical components of the displacement field in the silicon line are shown in Fig.~\ref{figSTI_displ}(a). They are retrieved by unwrapping the phases from Fig.~\ref{figSTI_result}(b) and substituting them into Eq.~\ref{phases}. The retrieved components of displacements are in very good agreement with the corresponding displacement field calculated by finite element modelling for the same semiconductor structures (\citeasnoun{Eberlein2008}). In Fig.~\ref{figSTI_displ}(b) the retrieved phases are depicted if the origin of the RSI is shifted to the second maxima of the RLP of the line ($\Delta \vec{h}^{004} = (\Delta h_x^{004}, \Delta h_z^{004}) = (0, -0.0145 \AA)$ in (\ref{phases})).

\section{Conclusion}

The analysis of the influence of additional constraints (\ref{constraint_phases2}) and (\ref{constraint_strains}) for coherent diffractive imaging of inhomogeneously strained crystals was performed. The suggested direct-space constraints limit the maximum magnitudes of the first and second order phase derivatives. One possible way of their implementation was suggested by modifying the hybrid input output algorithm to Eq.~\eqref{HIO_ap:a}--\eqref{HIO_ap:c}. The developed algorithm was tested on experimental diffraction data from periodic strained silicon lines.

The results from our reconstructions reveal the complexity of iterative reconstruction of the strained crystalline objects. Limiting the first order phase derivatives (\ref{constraint_phases2}) provided successful reconstruction only in case of subdivision of the full support into two regions (line and substrate) with individual bounds for every domain's constraints. This is additional \emph{a priori} knowledge about our system. However, convergence was still not reliable. 

The best results were observed when implementing only constraints to the second order phase derivatives (\ref{constraint_strains}). In this case, it was sufficient to consider the entire support as one unit: no subdivision was required. Convergence to the solution was observed to be much more stable. Residual reconstruction difficulties are most likely related to the artifacts of the measured experimental data.

Therefore, we could demonstrate a novel choice for additional \emph{a priori} knowledge to provide reconstruction of experimentally measured coherent diffractive imaging data. 

\ack{ATMEL company is gratefully acknowledged for financial support and sample supply. We especially thank Pascal Rohr and Romain Coppard from ATMEL. Michel Eberlein (ATMEL and IM2NP) performed his PhD work on stresses induced by STI. He performed the Finite Element Modelling. A. A. M and O. T. are grateful to Marc Gailhanou for his support during the experiment and for useful discussions. ESRF (Beamline BM32) is acknowledged for providing beam time. This work was supported by the program PNI of the Helmholtz Association.}

% \referencelist[biblio_CDI]% Produces the bibliography via BibTeX.

%\bibliographystyle{iucr}
%\bibliography{biblio_CDI}% Produces the bibliography via BibTeX.

\end{document}